\documentstyle[aps,prl,twocolumn,epsf,floats]{revtex}
\draft
\begin{document}
\title{Theory of internal transitions of charged excitons \\
  in quantum wells in magnetic fields}
\author{A. B. Dzyubenko}
\address{Institut f\"{u}r Theoretische Physik, J.W. Goethe-Universit\"{a}t,
         60054 Frankfurt,  Germany \\
General Physics Institute, RAS, Moscow 117942, Russia}
\author{A. Yu. Sivachenko}
\address{The Weizmann Institute of Science, Rehovot 76100, Israel}
\date{\today}
\maketitle
\begin{abstract}
For charged semiconductor complexes in magnetic fields $B$,
we discuss an exact classification of states, which is based
on magnetic translations. In this scheme, in addition to
the total orbital angular momentum projection $M_z$ and electron
and hole spins $S_e$, $S_h$, a new exact quantum
number appears. This oscillator quantum number, $k$,
is related physically to the
center of the cyclotron motion of the complex as a whole.
In the dipole approximation $k$ is strictly conserved
in magneto-optical transitions.
We discuss implications of this new exact selection rule for
internal intraband magneto-optical transitions of
charged excitons $X^-$ in quantum wells in $B$.
\end{abstract}

Recently, there has been considerable experimental and theoretical
interest in charged excitons $X^-$ and $X^+$ in magnetic fields $B$
in 2D systems.
Experimentally, magneto-optical interband \cite{inter}
transitions of charged excitons have been studied extensively.
Theoretically, the binding of charged excitons $X^-$ has been
considered in quantum dots \cite{Hawr},
in a strictly-2D system in the high-magnetic field limit \cite{AHM},
and in realistic quantum wells at finite $B$ \cite{Whit97}.
In all these theoretical works on charged excitons,
the existing exact symmetry --- magnetic translations ---
has not been identified.
The aim of the present theoretical work is to describe in some
detail this symmetry and its manifestations in intraband
magneto-optical transitions (see also \cite{Dz&S_pr}).
Experimental evidences for internal $X^-$ singlet and triplet transitions
have been reported very recently in \cite{EP2DS}, where also
comparison with quantitative calculations is presented.

We consider a system of interacting particles of charges $e_j$
in a magnetic field ${\bf B}=(0,0,B)$ described by the Hamiltonian
\begin{equation}
                \label{H} 
   H = \sum_j \frac{\hat{\bbox{\pi}}_j^2}{2m_j}
          + \case{1}{2} \sum_{i \ne j} U_{i j}({\bf r}_i-{\bf r}_j) \, ,
\end{equation}
here
$\hat{\bbox{\pi}}_j = -i\hbar \bbox{\nabla }_j -
\frac{e_j}{c} {\bf A}({\bf r}_j)$
is the kinematic momentum operator of the $j$-th particle in ${\bf B}$ and
$U_{ij}$ are the potentials of interactions that can be rather arbitrary.
Dynamical symmetries of (\ref{H}) are the following.
In the symmetric gauge ${\bf A} = \frac12 {\bf B} \times {\bf r}$,
there is the axial symmetry about the $z$-axis
$[H, \hat{L}_z]=0$, where
$\hat{L}_z=\sum_j ({\bf r}_j \times -i\hbar\bbox{\nabla }_j)_z$.
Therefore, the total angular momentum projection $M_z$,
an eigenvalue of $\hat{L}_z$, is a good quantum number.
In a uniform ${\bf B}$, the Hamiltonian (\ref{H}) is also invariant under
a group of magnetic translations whose generators are the components
of the operator $\hat{\bf K} = \sum_{j} \hat{\bf K}_j$,
where $\hat{\bf K}_j =
\hat{\bbox{\pi }}_j - \frac{e_j}{c} {\bf r}_j \times {\bf B}$
(see, e.g., \cite{Simon}).
$\hat{\bf K}$ is the exact integral of the motion: $[H, \hat{\bf K}]=0$.
The components of $\hat{\bf K}$ and
$\hat{\bbox{\pi}} = \sum_j \hat{\bbox{\pi}}_j $ commute in
${\bf B}$ as
\begin{equation}
        \label{comK}
 [\hat{K}_x, \hat{K}_y] = - [\hat{\pi}_x, \hat{\pi}_y] =
    - i \frac{\hbar B}{c} Q \quad , \quad
 \quad Q \equiv \sum_j e_j  \, ,
\end{equation}
while $[\hat{K}_p, \hat{\pi}_q]= 0$, $p,q=x,y$.
For neutral complexes (atoms, excitons, biexcitons) $Q=0$,
and classification of states in $B$ are due to the continuous two-component
vector --- the 2D magnetic momentum ${\bf K}= (K_x,K_y)$.
For charged systems the components of $\hat{\bf K}$
cannot be observed simultaneously.
This determines the macroscopic
Landau degeneracy of exact eigenstates of (\ref{H}).
For a dimensionless operator
$\hat{{\bf k}} = \sqrt{c/2 \hbar B |Q|} \, \hat{\bf K}$
we have $[\hat{k}_x, \hat{k}_y]=-iQ/|Q|$. Therefore,
$\hat{k}_{\pm}= (\hat{k}_x  \pm i \hat{k}_y)/\sqrt{2}$
are Bose raising and lowering ladder operators:
$[\hat{k}_{+}, \hat{k}_{-}]=-Q/|Q|$.  It follows then that
$\hat{{\bf k}}^2 =
\hat{k}_{+} \hat{k}_{-} + \hat{k}_{-} \hat{k}_{+}$
has discrete oscillator eigenvalues $2k+1$, $k=0, 1, \ldots$.
Since $[\hat{{\bf k}}^2, H]=0$ and $[\hat{{\bf k}}^2, \hat{L}_z]=0$,
the exact charged eigenstates of (\ref{H}), in addition
to the electron $S_e$ and hole $S_h$ spin quantum numbers,
can be simultaneously labeled
by the discrete quantum numbers $k$ and $M_z$.
The labelling therefore is $|k M_z S_e S_h \nu \rangle$.
Here $\nu$ is the ``principal'' quantum number, which
can be discrete (bound states) or continuous (unbound states forming
a continuum) \cite{Dz&S_pr}.
The $k=0$ states are {\em Parent States\/} (PS's)
within a degenerate manifold.
All other {\em Daughter\/} states in each $\nu$-th family
are generated out of the PS iteratively: for $Q<0$
$|k,M_z-k,S_e S_h \nu \rangle =
(\hat{k}_{-})^k |0, M_z, S_e S_h \nu \rangle / \sqrt{k!}$.
\begin{figure}[t]
\epsfxsize=3.1in
\epsffile{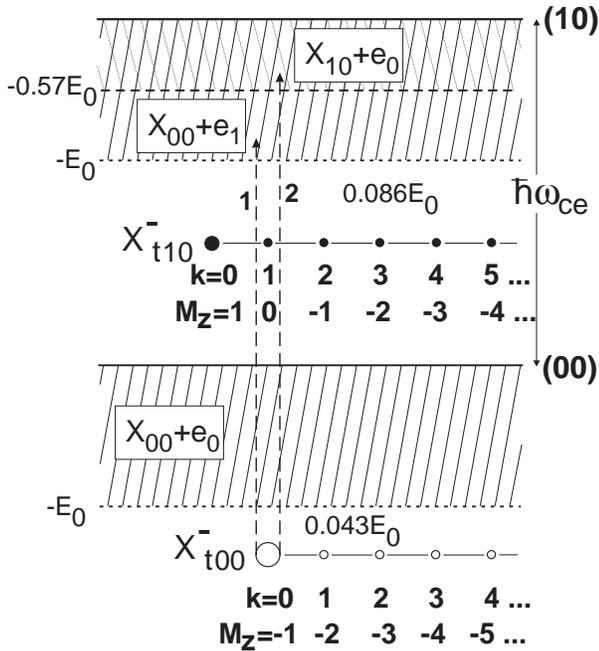}
\vspace*{5ex}
\caption{Schematic drawing of bound and scattering electron triplet
$2e$--$h$ states in the lowest LL's ($N_eN_h$)=(00), (10).
The states are labeled by the total angular momentum
projection $M_z$ and oscillator quantum number $k$.
Separation between the 2D $e$- and $h$-layers $d=0$ (see inset to Fig.\,2).
The energy $E_0= \protect\sqrt{\pi/2} \, e^2/\epsilon l_B$
parametrizes the 2D system in the limit of high $B$.
Large (small) dots correspond to the bound parent $k=0$ (daughter
$k=1, 2, \ldots$) $X^-$ states; see text for further explanations.}
                \label{fig1}
\end{figure}

Let us discuss now magneto-optical transitions of charged complexes.
In the dipole approximation the photon momentum is negligibly small.
Therefore, the quantum number $k$ should be conserved in
intra- and interband magneto-optical transitions.
For interband transitions in a translationally-invariant
system with a simple valence band this leads to a striking
result \cite{Dz&S_pr} that the ground triplet $X^-_t$ ($S_e=1$)
state is dark
in photoluminescence at {\em finite\/} magnetic fields $B$ and in
a quasi-2D system with a possible $e$--$h$ asymmetry (cf.\ \cite{AHM}).
Also, conservation of $k$ prohibits shake-up processes
of an isolated singlet $X^-_s$ ($S_e=0$) state in $B$.
In this work we concentrate on internal intraband
transitions of charged complexes in a magnetic field.
In these transitions conservation of $k$ follows from the commutativity
$[\hat{V}^{\pm}, \hat{\bf K}]=0$,
where
$\hat{V}^{\pm} \sim  \sum_{j} e_j \pi_{j}^{\pm}/m_j$,
is the Hamiltonian of the interaction with the
radiation of polarization $\sigma^{\pm}$
and $\pi_j^{\pm} = \pi_{jx} \pm i \pi_{jy}$.
Other selection rules in this case are conservation of spins $S_e$,
$S_h$ and $\Delta M_z= \pm 1$ in the $\sigma^{\pm}$ polarization
for the envelope function.
Conservation of $k$ constitutes a new exact selection rule.
It leads, in particular, to some striking consequences
for bound-to-bound transitions from the triplet $X_t$ ground state.

To show this we shall consider the system of strictly-2D $e$- and $h$-layers
separated by a distance $d$ (see inset to Fig.\,2)
in the limit of high magnetic fields.
To understand internal $X^-$ transitions, it is necessary to consider
the eigenstates associated with higher Landau levels (LL's).
To this end we use expansion in free LL's,
which has been described in some detail elsewhere \cite{Dz_PLA,Dz&S_pr}.
The calculated three-particle $2e$--$h$ eigenspectra
(electrons in the triplet state) in the two lowest LL's are shown
for $d=0$ in Fig.\,1.
\begin{figure}[t]
\epsfxsize=3.4in
\epsffile{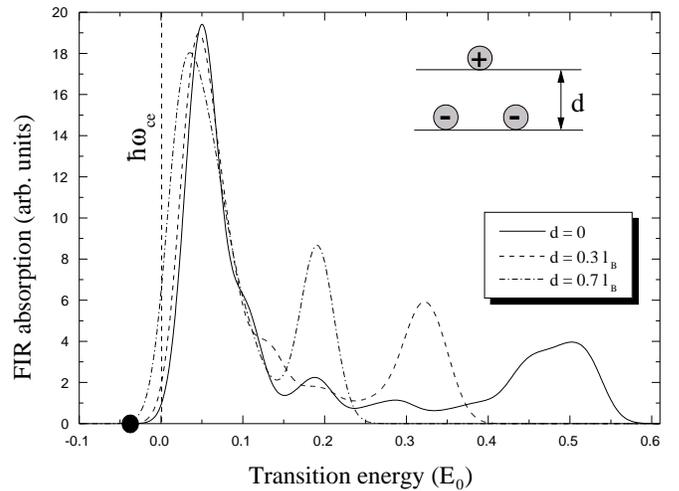}
\vspace*{5ex}
\caption{Energies and dipole matrix elements of the inter-LL transitions
from the ground $X^-_{t00}$ state in the high-field limit.
Shown are the cases of three different separations
$d=0$, $0.3l_B$, and $0.7l_B$
between the strictly-2D $e$- and $h$-layers (see inset).
A filled dot shows the position of the forbidden bound-to-bound
$X^-_{t00} \protect\rightarrow X^-_{t10}$ transition for $d=0$.}
\label{fig2}
\end{figure}
Generally, the eigenspectra  associated with each LL
consist of bands of {\em finite\/} width
$\sim E_0= \sqrt{\pi/2} \, e^2/\epsilon l_B$,
where $l_B=(\hbar c/e B)^{1/2}$.
The states within each such band form a {\em continuum\/}
corresponding to the extended motion of a neutral
magnetoexciton (MX) as a whole with the second electron
in a scattering state.
As an example, the continuum in the lowest ($N_eN_h$)=(00) LL
consists of the MX band of width $E_0$ extending down in energy
from the free $(00)$ LL\@.
This corresponds to the $1s$ MX ($N_e=N_h=0$) \cite{L&L80}
plus a scattered electron in the zero LL, labeled $X_{00}+e_0$.
The structure of the continuum in the ($N_eN_h$)=(10) LL is
more complicated:
in addition to the $X_{00}+e_1$ band of the width $E_0$,
there is another MX band of width $0.574 E_0$ also extending down
in energy from the free ($N_eN_h$)=(10) LL\@. This
corresponds to the $2p^+$ exciton ($N_e=1$, $N_h=0$) \cite{L&L80}
plus a scattered electron in the $N_e=0$ LL, labeled $X_{10}+e_0$.
There are also bands (not shown in Fig.\ 1)
above each free LL originating from the bound
internal motion of two electrons in the absence of a hole \cite{Dz&S_pr}.
Internal transitions to such bands have extremely small
oscillator strengths and not discussed here.
Bound $X^-$ states (finite internal motions of all three particles)
lie {\em outside\/} the continua (Fig.\,1).
In the limit of high $B$ the only bound $X^-$ state in the zeroth LL
($N_eN_h$)=(00) is the $X^-$-triplet. There are no bound
$X^-$-singlet states \cite{AHM,Whit97}
in contrast to the $B = 0$ case.
The $X^-$-triplet binding energy in zero LL's
($N_eN_h$)=(00) is $0.043 E_0$ \cite{AHM,Whit97}.
In the next electron LL ($N_eN_h$)=(10)
there are no bound $X^-$-singlets, and only one
bound triplet state $X^-_{t10}$, lying below the lower edge of the MX band
\cite{Dz&S_pr}.
The $X^-_{t10}$ binding energy is $0.086 E_0$, twice that of the $X^-_{t00}$,
and similar to the stronger binding of the $D^-$-triplet
in the $N_e=1$ LL \cite{Dz_PLA}.

We focus here on internal transitions in the $\sigma^+$ polarization
governed by the usual selection rules:
spin conserved, $\Delta M_z= 1$.
In this case the $e$-CR--like inter-LL ($\Delta N_e = 1$)
transitions are strong and gain strength with $B$.
Both bound-to-bound $X^-_{t00} \rightarrow X^-_{t10}$
and photoionizing $X^-$ transitions are possible.
For the latter the final three-particle states in the $(10)$ LL belong
to the continuum (Fig.\,1), and calculations show that
the FIR absorption spectra reflect its rich structure \cite{Dz&S_pr}.
Transitions to the $X_{00}+e_1$ continuum are
dominated by a sharp onset at the edge (transition~1)
at an energy $\hbar\omega_{ce}$ plus
the $X^-_{t00}$ binding energy.
In addition, there is a broader and weaker peak corresponding to
the transition to the $X_{01}+e_0$ MX band, transition~2.
The latter may be thought of as the $1s\rightarrow 2p^+$
internal transition of the MX \cite{Dz97},
which is shifted and broadened by the presence of the second electron.
In accordance with this picture, it is visible from Fig.\,2
that with increasing separation $d$ between the
$e$- and $h$-layers (when the exciton binding and, thus, transition
energies are reduced and the $X$--$e$ interaction is effectively diminished),
the second peak is redshifted and sharpened.
Thus the $X^-$-triplet behaves physically in the
photoionizing bound-to-continuum transitions as an exciton
that very loosely binds an electron, and the two ``parts'' of the complex can
absorb the FIR photon, to some extent, independently.
The double-peak structure of the bound-to-continuum
transitions is a generic feature for transitions
from both the singlet and triplet ground $X^-$ states in
quasi-2D systems in strong $B$.
Such transitions in translationally invariant systems are
discussed theoretically in \cite{EP2DS},
where also experimental results for bound-to-continuum transitions are
reported and comparison between theory and experiment is made.

The inter-LL bound-to-bound transition,
$X^-_{t00} \rightarrow X^-_{t10}$, has a very specific
spectral position: since the final state is more stronger
bound, it lies {\it below\/} the $e$-CR energy
$\hbar\omega_{ce}=\hbar eB/m_e c$.
However, it has exactly {\em zero\/} oscillator strength,
a manifestation of the magnetic translational invariance:
the two selection rules -- conservation of $k$ and $\Delta M_z =1$
cannot be satisfied simultaneously.
Indeed, e.g., the $X_{t00}$ PS (with $k=0$) has $M_z=-1$, while
the $X_{t10}$  PS has $M_z=1$, so that the usual selection rule
$\Delta M_z= 1$ cannot be satisfied.
Localization of charged excitons breaks translational
invariance and relaxes the $k$-conservation rule.
As a result, the bound-to-bound
$X^-_{t00} \protect\rightarrow X^-_{t10}$ transition,
which is prohibited in translationally-invariant systems,
develops {\em below\/} the $e$-CR \cite{Dz&S_pr}. Such a peak
is a tell-tale mark of localization of charged triplet excitons.

In conclusion, we have studied the exact symmetry  ---
magnetic translations ---  for charged excitons in $B$ and
established its consequences for intraband
magneto-optical transitions. In particular, we have shown
that in translationally invariant quasi-2D system with a simple valence
band the bound-to-bound transition from the triplet ground
state $X_t^-$ to the next electron Landau level
is prohibited. In the presence of translationally-breaking
effects (disorder, impurities etc.) the intraband
bound-to-bound triplet transition develops below the
electron cyclotron resonance.
This suggests a method of studying localization of charged
excitons.

ABD is grateful to the Alexander von Humboldt Foundation
for research support.


\end{document}